\documentclass[traditabstract]{aa}
\usepackage{amsmath}
\usepackage{txfonts}
\usepackage{graphicx}
\usepackage{natbib}
\bibpunct{(}{)}{;}{a}{}{,}

\begin{document}

\title{The generalised Lomb-Scargle periodogram}

\subtitle{A new formalism for the floating-mean and Keplerian periodograms}

\author{M. Zechmeister\inst{1}\and M. K\"{u}rster\inst{1}}

\institute{Max-Planck-Institut f\"{u}r Astronomie, K\"{o}nigstuhl 17, 69117 Heidelberg, Germany\\
\email{zechmeister@mpia.de}
}

\date{Received / Accepted}

\abstract{The Lomb-Scargle periodogram is a common tool in the frequency analysis of unequally spaced data equivalent to least-squares fitting of sine waves. We give an analytic solution for the generalisation to a full sine wave fit, including an offset and weights ($\chi^{2}$ fitting). Compared to the Lomb-Scargle periodogram, the generalisation is superior as it provides more accurate frequencies, is less susceptible to aliasing, and gives a much better determination of the spectral intensity. Only a few modifications are required for the computation and the computational effort is similar. Our approach brings together several related methods that can be found in the literature, viz. the date-compensated discrete Fourier transform, the floating-mean periodogram, and the ``spectral significance'' estimator used in the SigSpec program, for which we point out some equivalences. Furthermore, we present an algorithm that implements this generalisation for the evaluation of the Keplerian periodogram that searches for the period of the best-fitting Keplerian orbit to radial velocity data. The systematic and non-random algorithm is capable of detecting eccentric orbits, which is demonstrated by two examples and can be a useful tool in searches for the orbital periods of exoplanets.}


\maketitle

\section{Introduction}

The Lomb-Scargle periodogram \citep{Scargle82} is a widely used tool in period searches and frequency analysis of time series. It is equivalent to fitting sine waves of the form $y=a\cos\omega t+b\sin\omega t$. While standard fitting procedures require the solution of a set of linear equations for each sampled frequency, the Lomb-Scargle method provides an analytic solution and is therefore both convenient to use and efficient. The equation for the periodogram was given by \citet{Barning63}, and also \citet{Lomb76} and \citet{Scargle82}, who furthermore investigated its statistical behaviour, especially the statistical significance of the detection of a signal. For a time series ($t_{i}$, $y_{i}$) with zero mean ($\overline{y}=0$), the Lomb-Scargle periodogram is defined as (normalisation from \citealt{Lomb76}):
\begin{align}
\hat{p}(\omega) & =\frac{1}{\hat{YY}}\left[\frac{\hat{YC}_{\hat{\tau}}^{2}}{\hat{CC}_{\hat{\tau}}}+\frac{\hat{YS}_{\hat{\tau}}^{2}}{\hat{SS}_{\hat{\tau}}}\right]\label{eq:LS}\\
 & =\frac{1}{\sum_{i}y_{i}^{2}}\left\{ \frac{\left[\sum_{i}y_{i}\cos\omega(t_{i}-\hat{\tau})\right]^{2}}{\sum_{i}\cos^{2}\omega(t_{i}-\hat{\tau})}+\frac{\left[\sum_{i}y_{i}\sin\omega(t_{i}-\hat{\tau})\right]^{2}}{\sum_{i}\sin^{2}\omega(t_{i}-\hat{\tau})}\right\}
\end{align}
where the hats are used in this paper to symbolise the classical expressions. The parameter $\hat{\tau}$ is calculated via
\begin{equation}
\tan2\omega\hat{\tau}=\frac{\sum_{i}\sin2\omega t_{i}}{\sum_{i}\cos2\omega t_{i}}.\label{eq:LStau}
\end{equation}

However, there are two shortcomings. First, the Lomb-Scargle periodogram does not take the measurement errors into account. This was solved by introducing weighted sums by \citet{Gilliland87} and \citet{Irwin89} (equivalent to the generalisation to a $\chi^{2}$ fit). Second, for the analysis the mean of the data was subtracted, which assumes that the mean of the data and the mean of the fitted sine function are the same. One can overcome this assumption with the introduction of an offset $c$, resulting in a further generalisation of this periodogram to the equivalent of weighted full sine wave fitting; i.e., $y=a\cos\omega t+b\sin\omega t+c$. \citet{Cumming99}, who called this generalisation ``floating-mean periodogram'', argue that this approach is superior: \emph{``... the Lomb-Scargle periodogram fails to account for statistical fluctuations in
the mean of a sampled sinusoid, making it non-robust when the number of observations is small, the sampling is uneven, or for periods comparable to or greater than the duration of the observations.''} These authors provided a formal definition and also a sophisticated statistical treatment, but do not use an analytical solution for the computation of this periodogram.

Basically, analytical formulae for a full sine, least-squares spectrum have already been given by \citet{Ferraz81}, calling this date-compensated discrete Fourier transform (DCDFT). We prefer to adopt a notation closely related to the Lomb-Scargle periodogram calling it the generalised Lomb-Scargle periodogram (GLS). \citet{Shrager01} tries for such an approach but did not generalise the parameter $\hat{\tau}$ in Eq.~(\ref{eq:LStau}). Moreover, our generalised equations, which are derived in the following (Sect.~\ref{sec:GLS}), have a comparable symmetry to the classical ones and also allow us to point out equivalences to the ``spectral significance'' estimator used in the SigSpec program by \citet{Reegen07} (Sect.~\ref{sec:SigSpec}).

\section{\label{sec:GLS}The generalised Lomb-Scargle periodogram (GLS)}

The analytic solution for the generalised Lomb-Scargle periodogram can be obtained in a straightforward manner in the same way as outlined in \citet{Lomb76}. Let $y_{i}$ be the $N$ measurements of a time series at time $t_{i}$ and with errors $\sigma_{i}$. Fitting a full sine function (i.e. including an offset $c$):
\[
y(t)=a\cos\omega t+b\sin\omega t+c\]
at given frequency $\omega$ (or period $P=\frac{2\pi}{\omega}$) means to minimise the squared difference between the data $y_{i}$ and the model function $y(t)$:
\[
\chi^{2}=\sum_{i=1}^{N}\frac{[y_{i}-y(t_{i})]^{2}}{\sigma_{i}^{2}}=W\sum w_{i}[y_{i}-y(t_{i})]^{2}\]
where
\begin{align*}
w_{i}=\frac{1}{W}\frac{1}{\sigma_{i}^{2}} & \qquad\left(W=\sum\frac{1}{\sigma_{i}^{2}}\qquad\sum w_{i}=1\right)
\end{align*}
are the normalised weights\footnote{For clarity the bounds of the summation are suppressed in the following notation. They are always the same ($i=1,2,...,N$).}. Minimisation leads to a system of (three) linear equations whose solution is derived in detail in Appendix \ref{sub:Deriv_Periodogram}. Furthermore, it is shown in \ref{sub:Deriv_Periodogram} that the relative $\chi^{2}$-reduction $p(\omega)$ as a function of frequency $\omega$ and normalised to unity by $\chi_{0}^{2}$ (the $\chi^{2}$ for the weighted mean) can be written as:
\begin{align}
p(\omega) & =\frac{\chi_{0}^{2}-\chi^{2}(\omega)}{\chi_{0}^{2}}\label{eq:PeriodogramDef}\\
p(\omega) & =\frac{1}{YY\cdot D}\left[SS\cdot YC^{2}+CC\cdot YS^{2}-2CS\cdot YC\cdot YS\right]\label{eq:Periodogram}
\end{align}
with:
\begin{align}
D(\omega) & =CC\cdot SS-CS^{2}
\end{align}
and the following abbreviations for the sums:
\begin{alignat}{2}
 & Y & \,=\, & \sum w_{i}y_{i}\label{eq:Y}\\
 & C & =\, & \sum w_{i}\cos\omega t_{i}\\
 & S & =\, & \sum w_{i}\sin\omega t_{i}
\end{alignat}
\begin{alignat}{5}
 & YY & \,=\, & \hat{YY} & - & Y\cdot Y & \qquad\qquad & \hat{YY} & \,=\, & \sum w_{i}y_{i}^{2}\label{eq:YY}\\
 & YC(\omega) & =\, & \hat{YC} & - & Y\cdot C &  & \hat{YC} & =\, & \sum w_{i}y_{i}\cos\omega t_{i}\label{eq:YC}\\
 & YS(\omega) & =\, & \hat{YS} & - & Y\cdot S &  & \hat{YS} & =\, & \sum w_{i}y_{i}\sin\omega t_{i}\label{eq:YS}\\
 & CC(\omega) & =\, & \hat{CC} & - & C\cdot C &  & \hat{CC} & =\, & \sum w_{i}\cos^{2}\omega t_{i}\label{eq:CC}\\
 & SS(\omega) & =\, & \hat{SS} & - & S\cdot S &  & \hat{SS} & =\, & \sum w_{i}\sin^{2}\omega t_{i}\label{eq:SS}\\
 & CS(\omega) & =\, & \hat{CS} & - & C\cdot S &  & \hat{CS} & =\, & \sum w_{i}\cos\omega t_{i}\sin\omega t_{i}\label{eq:CS}
\end{alignat}
Note that sums with hats correspond to the classical sums. $W\cdot YY\equiv\chi_{0}^{2}$ is simply the weighted sum of squared deviations from the weighted mean. The mixed sums can also be written as a weighted covariance $Cov_{x,y}=\sum w_{i}x_{i}y_{i}-X\cdot Y/W=E(x\cdot y)-WE(x)E(y)$ where $E$ is the expectation value, e.g. $YS=Cov_{y,\sin\omega t}$.

With the weighted mean given by $\overline{y}=\sum w_{i}y_{i}=Y$ Eqs.~(\ref{eq:YY})-(\ref{eq:YS}) can also be written as:
\begin{alignat}{2}
 & YY & = & \sum w_{i}(y_{i}-\overline{y})^{2}\label{eq:YYal}\\
 & YC(\omega) & = & \sum w_{i}(y_{i}-\overline{y})\cos\omega t_{i}\label{eq:YCal}\\
 & YS(\omega) & = & \sum w_{i}(y_{i}-\overline{y})\sin\omega t_{i}.\label{eq:YSal}
\end{alignat}
So the sums $YC$ and $YS$ use the weighted mean subtracted data and are calculated in the same way as for the Lomb-Scargle periodogram (but with weights).

The generalised Lomb-Scargle periodogram $p(\omega)$ in Eq.~(\ref{eq:PeriodogramDef}) is normalised to unity and therefore in the range of $0\le p\le1$, with $p=0$ indicating no improvement of the fit and $p=1$ a ``perfect'' fit (100\% reduction of $\chi^{2}$ or $\chi^{2}=0$).

As the full sine fit is time-translation invariant, there is also the possibility to introduce an arbitrary time reference point $\tau$ ($t_{i}\rightarrow t_{i}-\tau$; now, e.g. $CC=\sum w_{i}\cos^{2}\omega(t_{i}-\tau)-\left(\sum w_{i}\cos\omega(t_{i}-\tau)\right)^{2}$), which will not affect the $\chi^{2}$ of the fit. If this parameter $\tau$ is chosen as
\begin{align}
\tan2\omega\tau & =\frac{2CS}{CC-SS}\label{eq:tau}\\
 & =\frac{\sum w_{i}\sin2\omega t_{i}-2\sum w_{i}\cos\omega t_{i}\sum w_{i}\sin\omega t_{i}}{\sum w_{i}\cos2\omega t_{i}-\left[\left(\sum w_{i}\cos\omega t_{i}\right)^{2}-\left(\sum w_{i}\sin\omega t_{i}\right)^{2}\right]}\nonumber
\end{align}
the interaction term in Eq.~(\ref{eq:Periodogram}) disappears, $CS_{\tau}=\sum w_{i}\cos\omega(t_{i}-\tau)\sin\omega(t_{i}-\tau)-\sum w_{i}\cos\omega(t_{i}-\tau)\sum w_{i}\sin\omega(t_{i}-\tau)=0$ (proof in Appendix \ref{sub:Proof_tau}) and in this case we append the index $\tau$ to the time dependent sums. The parameter $\tau(\omega)$ is determined by the times $t_{i}$ and the measurement errors $\sigma_{i}$ for each frequency $\omega$. So when using $\tau$ as defined in Eq.~(\ref{eq:tau}) the periodogram in Eq.~(\ref{eq:Periodogram}) becomes
\begin{equation}
p(\omega)=\frac{1}{YY}\left[\frac{YC_{\tau}^{2}}{CC_{\tau}}+\frac{YS_{\tau}^{2}}{SS_{\tau}}\right].\label{eq:GLS}
\end{equation}
Note that Eq.~(\ref{eq:GLS}) has the same form as the Lomb-Scargle periodogram in Eq.~(\ref{eq:LS}) with the difference that the errors can be weighted (weights $w_{i}$ in all sums) and that there is an additional second term in $CC_{\tau}$, $SS_{\tau}$, $CS_{\tau}$ and $\tan2\omega\tau$ (Eqs.~(\ref{eq:CC})--(\ref{eq:CS}) and Eq.~(\ref{eq:tau}), respectively) which accounts for the floating mean.

The computational effort is similar as for the Lomb-Scargle periodogram. The incorporation of the offset $c$ requires only two additional sums for each frequency $\omega$ (namely $S=\sum w_{i}\sin\omega t_{i}$ and $C=\sum w_{i}\cos\omega t_{i}$ or $S_{\tau}$ and $C_{\tau}$ respectively). The effort is even weaker when using Eq.~(\ref{eq:Periodogram}) with keeping $CS$ instead of using Eq.~(\ref{eq:GLS}) with the parameter $\tau$ introduced via Eq.~(\ref{eq:tau}) which needs an extra preceding loop in the algorithm. If the errors are taken into account as weights, also the multiplication with $w_{i}$ must be done.

For fast computation of the trigonometric sums the algorithm of \citet{Press89} can be applied, which has advantages in the case of large data sets and/or many frequency steps. Another possibility are trigonometric recurrences\footnote{E.g. $\cos\omega_{k+1}t=\cos(\omega_{k}+\Delta\omega)t=\cos\omega_{k}t\cos\Delta\omega_{k}t-\sin\omega_{k}t\sin\Delta\omega t$ where $\Delta\omega$ is the frequency step.} as described in \citet{Press92}. Note also that the first sum in $SS$ can be expressed by $\hat{SS}=1-\hat{CC}$.

\section{Normalisation and False-Alarm probability (FAP)}

There were several discussions in the literature on how to normalise the periodogram. For the detailed discussion we refer to the key papers by \citet{Scargle82}, \citet{Horne86}, \citet{Koen90} and \citet{Cumming99}. The normalisation becomes important for estimations of the false-alarm probability of a signal by means of an analytic expression. \citet{Lomb76} showed that if data are Gaussian noise, the terms $\hat{YC}^{2}/\hat{CC}$ and $\hat{YS}^{2}/\hat{SS}$ in Eq.~(\ref{eq:LS}) are $\chi^{2}$-distributed and therefore the sum of both (which is $\propto p$) is $\chi^{2}$-distributed with two degrees of freedom. This holds for the generalisation in Eq.~(\ref{eq:GLS}) and also becomes clear from the definition of the periodogram in Eq.~(\ref{eq:PeriodogramDef}) $p(\omega)=\frac{\chi_{0}^{2}-\chi^{2}(\omega)}{\chi_{0}^{2}}$ where for Gaussian noise the difference in the numerator $\chi_{0}^{2}-\chi^{2}(\omega)$ is $\chi^{2}$-distributed with $\nu=(N-1)-(N-3)=2$ degrees of freedom.

The $p(\omega)$ can be compared with a known noise level $p_{n}$ (expected from the \emph{a priori} known noise variance or population variance) and the normalisation of $p(\omega)$ to $p_{n}$
\begin{equation}
P_{n}=\frac{p(\omega)}{p_{n}}\label{eq:NormScargle}
\end{equation}
can be considered as a signal to noise ratio \citep{Scargle82}. However, this noise level is often not known.

Alternatively, the noise level may be estimated for Gaussian noise from Eq.~(\ref{eq:PeriodogramDef}) to be $p_{n}=\frac{2}{N-1}$ which leads to:
\begin{equation}
P=\frac{N-1}{2}p(\omega)\label{eq:NormHorne}
\end{equation}
and is the analogon to the normalisation in \citet{Horne86}\footnote{These authors called it the normalization with the sample variance $\sigma_{0}^{2}$. Note that $p(\omega)$ is already normalized with $\chi_{0}^{2}$. For the unweighted case ($w_{i}=\frac{1}{N}$, $\sigma_{0}^{2}=\frac{N}{N-1}YY$) one can write Eq.~(\ref{eq:NormHorne}) with Eq.~(\ref{eq:GLS}) as $P(\omega)=\frac{1}{2}\frac{N}{\sigma_{0}^{2}}\left[\frac{YC^{2}}{CC}+\frac{YS^{2}}{SS}\right]$.}. So if the data are noise, $P=1$ is the expected value. If the data contains a signal, $P\gg1$ is expected at the signal frequency. However, this power is restricted to $0\le P\le\frac{N-1}{2}$.

But if the data contains a signal or if errors are under- or overestimated or if intrinsic variability is present, then $p_{n}=\frac{2}{N-1}$ may not be a good uncorrelated estimator for the noise level. \citet{Cumming99} suggested to estimate the noise level \emph{a posteriori} with the residuals of the best fit and normalised the periodogram as
\begin{align}
z(\omega) & =\frac{N-3}{2}\frac{\chi_{0}^{2}-\chi^{2}(\omega)}{\chi_{\mathrm{best}}^{2}}=\frac{N-3}{2}\frac{p(\omega)}{1-p_{\mathrm{best}}}\label{eq:NormCumming}
\end{align}
where the index ``best'' denotes the corresponding values of the best fit ($p_{\mathrm{best}}=p(\omega_{\mathrm{best}})$).

Statistical fluctuations or a present signal may lead to a larger periodogram power. To test for the significance of such a peak in the periodogram the probability is assessed that this power can arise purely from noise. \citet{Cumming99} clarified that the different normalisations result in different probability functions which are summarised in Table~\ref{Tab:Prob}.
Note that the last two probability values are the same for the best fit ($z_{0}=z_{\mathrm{best}}$):
\begin{align*}
\mathrm{Prob}(z>z_{\mathrm{best}}) & =\left(1+\frac{2z_{\mathrm{best}}}{N-3}\right)^{-(N-3)/2}=\left(1+\frac{p_{\mathrm{best}}}{1-p_{\mathrm{best}}}\right)^{-(N-3)/2}\\
 & =\left(\frac{1}{1-p_{\mathrm{best}}}\right)^{-(N-3)/2}=\left(1-p_{\mathrm{best}}\right)^{(N-3)/2}\\
 & =\mathrm{Prob}(p>p_{\mathrm{best}})=\mathrm{Prob}(P>P_{\mathrm{best}}).
\end{align*}
Furthermore \citet{Baluev08} pointed out that the power definition
\[
Z(\omega)=\frac{N-2}{3}\ln\frac{\chi_{0}^{2}}{\chi^{2}(\omega)}\]
as a nonlinear (logarithmic) scale for $\chi^{2}$ has an exponential distribution (similiar to $P_{n}$)
\[
\mathrm{Prob}(Z>Z_{\mathrm{best}})=e^{-Z}=\left(\frac{\chi^{2}(\omega)}{\chi_{0}^{2}}\right)^{(N-3)/2}=\mathrm{Prob}(p>p_{\mathrm{best}}).\]

\begin{table}
\caption{\label{Tab:Prob}Probabilities that a periodogram power ($P_{n}$, $p$, $P$ or $z$) can exceed a given value ($P_{n,0}$, $p_{0}$, $P_{0}$ or $z_{0}$) for different normalizations (from \citealt{Cumming99}).}
\centering
\begin{tabular}{lll}
\hline
\hline Reference level & Range & Probability\\
\hline
population variance & $P_{n}\in[0,\infty)$    & $\mathrm{Prob}(P_{n}>P_{n,0})=\exp(-P_{n,0})$\\
sample variance     & $p\in[0,1]$             & $\mathrm{Prob}(p>p_{0})=\left(1-p_{0}\right)^{\frac{N-3}{2}}$\\
-- '' --            & $P\in[0,\frac{N-1}{2}]$ & $\mathrm{Prob}(P>P_{0})=\left(1-\frac{2P}{N-1}\right)^{\frac{N-3}{2}}$\\
residual variance   & $z\in[0,\infty)$        & $\mathrm{Prob}(z>z_{0})=\left(1+\frac{2z_{0}}{N-3}\right)^{-\frac{N-3}{2}}$\\
\hline
\end{tabular}
\end{table}

Since in period search with the periodogram we study a range of frequencies, we are also interested in the significance of one peak compared to the peaks at other frequencies rather than the significance of a single frequency. The false alarm probability (FAP) for the period search in a frequency range is given by
\begin{equation}
\mathrm{FAP}=1-[1-\mathrm{Prob}(z>z_{0})]^{M}\label{eq:FAP}
\end{equation}
where $M$ is the number of independent frequencies and may be estimated as the number of peaks in the periodogram. The width of one peak is $\delta f\approx\frac{1}{T}$ ($\approx$frequency resolution). So in the frequency range $\Delta f=f_{2}-f_{1}$ there are approximately $M=\frac{\Delta f}{\delta f}$ peaks and in the case $f_{1}\ll f_{2}$ one can write $M\approx Tf_{2}$ \citep{Cumming04}. Finally, for low FAP values the following handy approximation for Eq.~(\ref{eq:FAP}) is valid:
\begin{equation}
\mathrm{FAP}\approx M\cdot\mathrm{Prob}(z>z_{0})\quad\text{for}\quad\mathrm{FAP}\ll1.\label{eq:FAPapprox}
\end{equation}

Another possibility to access $M$ and the FAP are Monte Carlo or bootstrap simulations in order to determine how often a certain power level ($z_{0}$) is exceeded just by chance. Such numerical calculation of the FAP are much more time-consuming than the actual computation of the GLS.

\section{\label{sec:SigSpec}Equivalences between the GLS and SigSpec \citep{Reegen07}}

\citet{Reegen07} developed a method, called SigSpec, to determine the significance of a peak at a given frequency (spectral significance) in a discrete Fourier transformation (DFT) which includes a zero mean correction. We will recapitulate some points from his paper in an adapted and shortened way in order to show several equivalences and to disentangle different notations used by us and used by \citet{Reegen07}. For a detailed description we refer to the original paper. Briefly, approaching from Fourier theory \citet{Reegen07} defined the zero mean corrected Fourier coefficients\footnote{Here only the unweighted case is discussed ($w_{i}=\frac{1}{N}$). \citet{Reegen07} also gives a generalization to weighting.}
\begin{align*}
a_{\mathrm{ZM}}(\omega) & =\frac{1}{N}\sum y_{i}\cos\omega t_{i}-\frac{1}{N^{2}}\sum y_{i}\cdot\sum\cos\omega t_{i}\\
b_{\mathrm{ZM}}(\omega) & =\frac{1}{N}\sum y_{i}\sin\omega t_{i}-\frac{1}{N^{2}}\sum y_{i}\cdot\sum\sin\omega t_{i}
\end{align*}
which obviously correspond to $YC$ and $YS$ in Eqs.~(\ref{eq:YC}) and (\ref{eq:YS}). Their variances are given by
\[
\left\langle a_{\mathrm{ZM}}^{2}\right\rangle =\frac{\left\langle y^{2}\right\rangle }{N^{2}}\left[\sum\cos^{2}\omega t_{i}-\frac{1}{N}\left(\sum\cos\omega t_{i}\right)^{2}\right]\]
\[
\left\langle b_{\mathrm{ZM}}^{2}\right\rangle =\frac{\left\langle y^{2}\right\rangle }{N^{2}}\left[\sum\sin^{2}\omega t_{i}-\frac{1}{N}\left(\sum\sin\omega t_{i}\right)^{2}\right]\]
The precise value of these variances depends on the temporal sampling. These variances can be expressed as $\left\langle a_{\mathrm{ZM}}^{2}\right\rangle =\frac{\left\langle y^{2}\right\rangle }{N}CC$ and $\left\langle b_{\mathrm{ZM}}^{2}\right\rangle =\frac{\left\langle y^{2}\right\rangle }{N}SS$.

Consider now two independent Gaussian variables whose cumulative distribution function (CDF) is given by:
\[
\Phi(\alpha,\beta|\omega)=e^{-\frac{1}{2}\left(\frac{\alpha^{2}}{\left\langle \alpha^{2}\right\rangle }+\frac{\beta^{2}}{\left\langle \beta^{2}\right\rangle }\right)}.\]
A Gaussian distribution of the physical variable $y_{i}$ in the time domain will lead to Gaussian variables $a_{\mathrm{ZM}}$ and $b_{\mathrm{ZM}}$ which in general are correlated. A rotation of Fourier Space by phase $\theta_{0}$
\[
\tan2\theta_{0}=\frac{N\sum\sin2\omega t_{i}-2\sum\cos\omega t_{i}\sum\sin\omega t_{i}}{N\sum\cos2\omega t_{i}-(\sum\cos\omega t_{i})^{2}+(\sum\sin\omega t_{i})^{2}}\]
transforms $a_{\mathrm{ZM}}$ and $b_{\mathrm{ZM}}$ into uncorrelated coefficients $\alpha$ and $\beta$ with vanishing covariance. Indeed, $\omega\tau$ from Eq.~(\ref{eq:tau}) and $\theta_{0}$ have the same value, but $\tau$ is applied in the time domain, while $\theta_{0}$ is applied to the phase $\theta$ in the Fourier domain. It is only mentioned here that the resulting coefficients $2\alpha$ and $2\beta$ correspond to $YC_{\tau}$ and $YS_{\tau}$.

Finally, \citet{Reegen07} defines as the spectral significance $\mathrm{sig}(\alpha,\beta|\omega):=-\log\Phi(\alpha,\beta|\omega)$ and writes:
\begin{align}
\mathrm{sig}(a_{\mathrm{ZM}},b_{\mathrm{ZM}}|\omega) & =\frac{N\log e}{\left\langle y^{2}\right\rangle }\left[\left(\frac{a_{\mathrm{ZM}}\cos\theta_{0}+b_{\mathrm{ZM}}\sin\theta_{0}}{\alpha_{0}}\right)^{2}\right.\label{eq:Sig}\\
 & \quad\quad\quad\quad\left.+\left(\frac{a_{\mathrm{ZM}}\cos\theta_{0}-b_{\mathrm{ZM}}\sin\theta_{0}}{\beta_{0}}\right)^{2}\right]\nonumber
\end{align}
where
\begin{align*}
\alpha_{0} & :=\sqrt{2N\frac{\left\langle \alpha^{2}\right\rangle }{\left\langle y^{2}\right\rangle }}\\
 & =\sqrt{\frac{2}{N^{2}}\left\{ N\sum\cos^{2}(\omega t_{i}-\theta_{0})-\left[\sum\cos(\omega t_{i}-\theta_{0})\right]^{2}\right\} }
\end{align*}
\begin{align*}
\beta_{0} & :=\sqrt{2N\frac{\left\langle \beta^{2}\right\rangle }{\left\langle y^{2}\right\rangle }}\\
 & =\sqrt{\frac{2}{N^{2}}\left\{ N\sum\sin^{2}(\omega t_{i}-\theta_{0})-\left[\sum\sin(\omega t_{i}-\theta_{0})\right]^{2}\right\} }
\end{align*}
are called normalised semi-major and semi-minor axes. Note that $\alpha_{0}^{2}\sim2CC_{\tau}$ and $\beta_{0}^{2}\sim2SS_{\tau}$.

\citet{Reegen07} states that this gives as accurate frequencies as do least squares. However, from this derivation it is not clear that this is equivalent. But when comparing Eq.~(\ref{eq:Sig}) to Eq.~(\ref{eq:GLS}) with using $YC_{\tau}=YC\cos\omega\tau+YS\sin\omega\tau$ and $YS_{\tau}=YS\cos\omega\tau-YC\sin\omega\tau$:
\begin{align*}
p(\omega) & =\frac{1}{YY}\left[\frac{(YC\cos\omega\tau+YS\sin\omega\tau)^{2}}{CC_{\tau}}\right.\\
 & \quad\quad\quad\left.+\frac{(YS\cos\omega\tau-YC\sin\omega\tau)^{2}}{SS_{\tau}}\right]
\end{align*}
the equivalence of the GLS and the spectral significance estimator in SigSpec (and with it to least squares) is evident:
\[
\mathrm{sig}(a_{\mathrm{ZM}},b_{\mathrm{ZM}}|\omega)=\frac{YY\cdot N\log e}{2\left\langle y^{2}\right\rangle }\cdot p(\omega).\]
The two indicators differ only in a normalisation factor, which becomes $\frac{N-1}{2}\log e$, when $\left\langle y^{2}\right\rangle $ is estimated with the sample variance $\left\langle y_{i}^{2}\right\rangle =\frac{N}{N-1}YY$. Therefore, SigSpec gives accurate frequencies just like least squares. But note that the Fourier amplitude is given by the sum of the squared Fourier coefficient $A^{2}=a_{ZM}^{2}+b_{ZM}^{2}=4\alpha^{2}+4\beta^{2}\sim YC^{2}+YS^{2}=YC_{\tau}^{2}+YS_{\tau}^{2}$ while the least-squares fitting amplitude is $A^{2}=a^{2}+b^{2}=\frac{YC_{\tau}^{2}}{CC_{\tau}^{2}}+\frac{YS_{\tau}^{2}}{SS_{\tau}^{2}}$ (see \ref{sub:Deriv_Periodogram}).

The comparison with SigSpec offers another point of view. It shows how the GLS is associated to Fourier theory and how it can be derived from the DFT (discrete Fourier transform) when demanding certain statistical properties such as simple statistical behaviour, time-translation invariance \citep{Scargle82} and varying zero mean. The shown equivalences allow \emph{vice versa} to apply several of \citeauthor{Reegen07}'s \citeyearpar{Reegen07} conclusions to the GLS, e.g. that it is less susceptible to aliasing or that the time domain sampling is taken into account in the probability distribution.

\section{Application of the GLS to the Keplerian periodogram (Keplerian fits to radial velocity data)}

The search for the best-fitting sine function is a multidimensional $\chi^{2}$-minimisation problem with four parameters: period $P$, amplitude $A$, phase $\varphi$ and offset $c$ (or frequency $\omega=2\pi/P$, $a$, $b$ and $c$). At a given frequency $\omega$ the best-fitting parameters $A$, $\varphi$ and $c$ can be computed immediately by an analytic solution revealing the global optimum for this three dimensional parameter subspace. But involving the frequency leads to a lot of local optima (minima in $\chi^{2}$) as visualised by the numbers of maxima in the generalised Lomb-Scargle periodogram.
With stepping through frequency $\omega$ one can pick up the global optimum in the four dimensional parameter space. That is how period search with the periodogram works. Because an analytic solution (implemented in the GLS) can be employed partially, there is no need for stepping through all parameters to explore the whole parameter space for the global optimum.

This concept can be transferred to the Keplerian periodogram which can be applied to search stellar radial velocity data for periodic signals caused by orbiting companions and measured from spectroscopic Doppler shifts. The radial velocity curve becomes more non-sinusoidal for a more eccentric orbit and its shape depends on six orbital elements%
\footnote{ $K=\frac{2\pi}{P}\frac{a\sin i}{\sqrt{1-e^{2}}}$ with $a$ the semi-major axis of the stellar orbit and $i$ the inclination.}:
\begin{alignat*}{2}
 & \gamma & \qquad & \text{constant system radial velocity}\\
 & K & \qquad & \text{radial velocity amplitude}\\
 & \varpi & \qquad & \text{longitude of periastron}\\
 & e & \qquad & \text{eccentricity}\\
 & T_{0} & \qquad & \text{periastron passage}\\
 & P & \qquad & \text{period}.
\end{alignat*}

In comparison to the full sine fit there are two more parameters to deal with which complicates the period search. An approach to simplify the Keplerian orbit search is to use the GLS to look for a periodic signal and use it for an initial guess. But choosing the best-fitting sine period does not necessarily lead to the best-fitting Keplerian orbit. So for finding the global optimum the whole parameter space must be explored.

The Keplerian periodogram \citep{Cumming04}, just like the GLS, shows how good a trial period (frequency $\omega$) can model the observed radial velocity data and can be defined as ($\chi^{2}$-reduction):
\[
p_{\mathrm{Kep}}(\omega)=\frac{\chi_{0}^{2}-\chi_{\mathrm{Kep}}^{2}(\omega)}{\chi_{0}^{2}}.\]
Instead of the sine function, the function
\begin{equation}
RV(t)=\gamma+K[e\cos\varpi+\cos(\nu(t)+\varpi)]\label{eq:RV}
\end{equation}
which describes the radial reflex motion of a star due to the gravitational pull of a planet, serves as the model for the radial velocity curve. The time dependence is given by the true anomaly $\nu$ which furthermore depends on three orbital parameters ($\nu(t,P,e,T_{0})$). The relation between $\nu$ and time $t$ is:
\[
\tan\frac{\nu}{2}=\sqrt{\frac{1+e}{1-e}}\tan\frac{E}{2}\]
\begin{equation}
E-e\sin E=M=2\pi\frac{t-T_{0}}{P}\label{eq:KepEq}
\end{equation}
where $E$ and $M$ are called eccentric anomaly and mean anomaly, respectively\footnote{The following expressions are also used frequently: $\sin\nu=\frac{\sqrt{1-e^{2}}\sin E}{1-e\cos E}$ and $\cos\nu=\frac{\cos E-e}{1-e\cos E}$.}. Eq.~(\ref{eq:KepEq}), called Kepler's equation, is transcendent meaning that for a given time $t$ the eccentric anomaly $E$ cannot be calculated explicitly.

For the computation of a Keplerian periodogram $\chi^{2}$ is to be minimised with respect to five parameters at a fixed trial frequency $\omega$. Similar to the GLS there is no need for stepping through all parameters. With the substitutions $c=\gamma+Ke\cos\varpi$, $a=K\cos\varpi$ and $b=-K\sin\varpi$ Eq.~(\ref{eq:RV}) can be written as:
\[
RV(t)=c+a\cos\nu(t)+b\sin\nu(t)\]
and with respect to the parameters $a$, $b$ and $c$ the analytic solution can be employed as described in Sect.~\ref{sec:GLS} for known $\nu$ (instead of $\omega t$) . So for fixed $P$, $e$ and $T_{0}$ the true anomalies $\nu_{i}$ can be calculated and the GLS from Eq.~(\ref{eq:Periodogram}) (now using $\nu_{i}$ instead of $\omega t_{i}$) can be applied to compute the maximum $\chi^{2}$-reduction ($p(\omega)$), here called $p_{e,T_{0}}(\omega)$. Stepping through $e$ and $T_{0}$ yields the best Keplerian fit at frequency $\omega$:
\[
p_{\mathrm{Kep}}(\omega)=\max_{e,T_{0}}p_{e,T_{0}}(\omega)\]
as visualised in the Keplerian periodogram. Finally, with stepping through the frequency, like for the GLS, one will find the best-fitting Keplerian orbit having the overall maximum power:
\[
p_{\mathrm{Kep}}(\omega_{\mathrm{best}})=\max_{\omega}p_{\mathrm{Kep}}(\omega)\]

There exist a series of tools (or are under development) using genetic algorithms or Bayesian techniques for fast searches for the best Keplerian fit \citep{Ford07,Balan08}. The algorithm, presented in this section, is not further optimised for speed. But it works well, is easy to implement and is robust since in principle it cannot miss a peak if the 3 dimensional grid for $e$, $T_{0}$ and $\omega$ is sufficiently dense. A reliable algorithm is needed for the computation of the Keplerian periodogram which by definition yields the best fit at fixed frequency and no local $\chi^{2}$-minima. \citet{OToole07,OToole08} developed an algorithm called 2DKLS (two dimensional Kepler Lomb-Scargle) that works on grid of period and eccentricity and seems to be similar to ours. But the possibility to use partly an analytic solution or the need for stepping $T_{0}$ is not mentioned by these authors.

The effort to compute the Keplerian periodogram with the described algorithm is much stronger in comparison to the GLS. There are three additional loops: two loops for stepping through $e$ and $T_{0}$ and one for the iteration to solve Kepler's equation. Contrary to the GLS it is not possible to use recurrences or the fast computation of the trigonometric sums mentioned in Sect.~\ref{sec:GLS}.

However we would like to outline some possibilities for technical improvements for a faster search. The first concerns the grid size. We choose a regular grid for $e$, $T_{0}$ and $\omega$. While this is adequate for the frequency $\omega$, as we discuss later in this section, there might be a more appropriate $e$-$T_{0}$ grid, e.g. a less dense grid size for $T_{0}$ at lower eccentricities. A second possibility is to reduce the iterations for solving Kepler's equation by using the eccentric anomalies (or differentially corrected ones) as initial values for the next slightly different grid point. This can save several ten percent of computation time, in particular in dense grids and at high eccentricities. A third idea which might have a high potential for speed up is to combine our algorithm with other optimisation techniques (e.g. Levenberg-Marquardt instead of pure stepping) to optimise the remaining parameters in the function $p_{e,T_{0}}(\omega)$. A raw grid would provide the initial values and the optimisation technique would do the fine adjustment.

\begin{figure}
\includegraphics{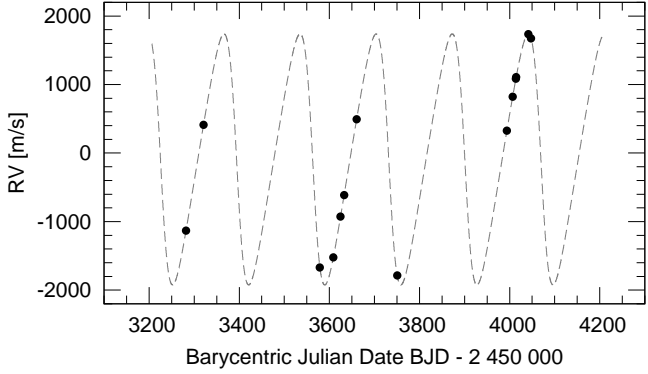}
\caption{\label{Fig:RVGJ1046}The radial velocity (RV) time series of the M dwarf GJ~1046. The solid line is the best Keplerian orbit fit ($P=168.8\,\mathrm{d}$, $e=0.28$).}
\end{figure}

\begin{figure}
\includegraphics{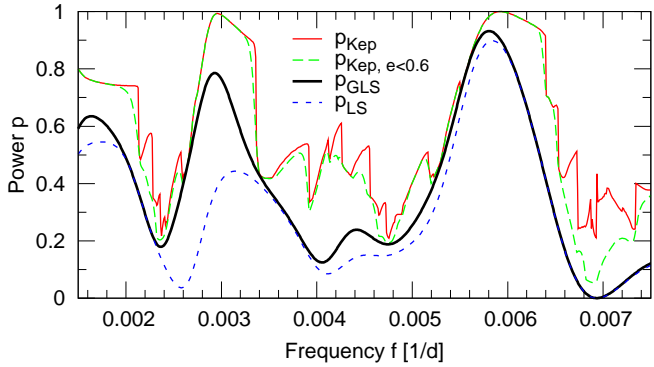}
\caption{\label{Fig:LSvsGLS}Comparison of the normalized Lomb-Scargle (LS), GLS and Keplerian periodograms for GJ~1046 ($f=\frac{1}{P}=\frac{\omega}{2\pi}$).}
\end{figure}

To give an example Fig.~\ref{Fig:RVGJ1046} shows RV data for the M dwarf GJ~1046 \citep{Kuerster08} along with the best-fitting Keplerian orbit ($P=168.8\,\mathrm{d}$, $e=0.28$). Figure~\ref{Fig:LSvsGLS} shows the Lomb-Scargle, GLS and Keplerian periodograms. Because a Keplerian orbit has more degrees of freedom it always has the highest $\chi^{2}$ reduction ($0\le p_{\mathrm{LS}}\le p_{\mathrm{GLS}}\le p_{\mathrm{Kep},e<0.6}\le p_{\mathrm{Kep}}\le1$).

As a comparison the Keplerian periodogram restricted to $e<0.6$ is also shown in Fig.~\ref{Fig:LSvsGLS}. At intervals where $p_{\mathrm{Kep}}$ exceeds $p_{e<0.6}$ the contribution is due to very eccentric orbits. Note that the Keplerian periodogram obtains more structure when the search is extended to more eccentric orbits. Therefore the evaluation of the Keplerian periodogram needs a higher frequency resolution (this effect can also be observed in \citealt{OToole08}). This is a consequence of the fact that more eccentric orbits are spikier and thus more sensitive to phase and frequency.

Other than \citet{OToole08} we plot the periodograms against frequency\footnote{for Fourier transforms this is common} to illustrate that the typical peak width $\delta f$ is frequency independent. Thus equidistant frequency steps ($\mathrm{d}f<\delta f$) yield a uniform sampling of each peak and are the most economic way to compute the periodogram rather than e.g. logarithmic period steps (leading to oversampling at long periods: $\mathrm{d}f=\mathrm{d}\frac{1}{P}=-\frac{1}{P^{2}}\mathrm{d}P=-\frac{1}{P}\mathrm{d}\ln P$)
as used by \citet{OToole08}. Still the periodograms can be plotted against a logarithmic period scale as e.g. sometimes preferred to present a period search for exoplanets.

\begin{figure}
\includegraphics{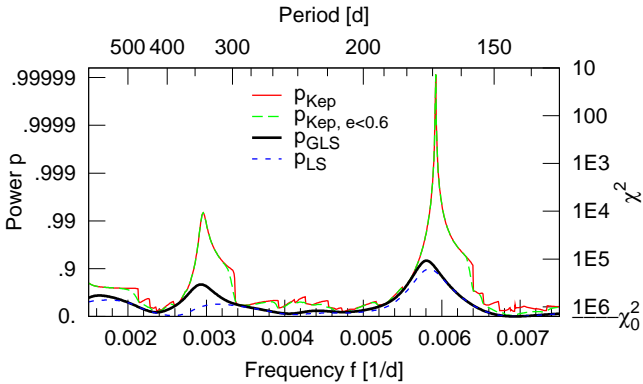}
\caption{\label{Fig:LSvsGLSlogscale}The same periodograms as in Fig.~\ref{Fig:LSvsGLS} with a quasi-logarithmic scale for $p$ and a logarithmic scale for $\chi^{2}$ (axis to the right).}
\end{figure}

\begin{figure}
\includegraphics{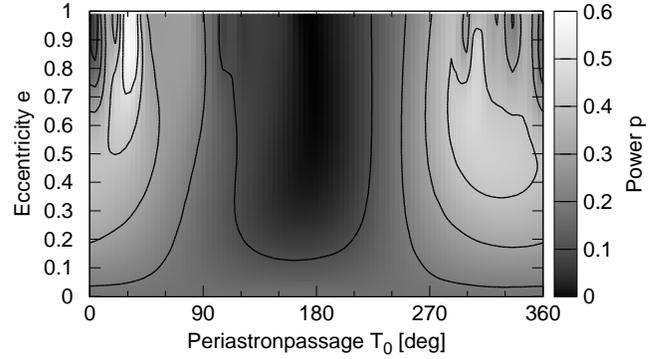}
\caption{\label{Fig:powermap} Power map ($p_{e,T_{0}}$) for $e$ and $T_{0}$ at the arbitrary fixed frequency $f=0.00422\,\mathrm{d}^{-1}$.
The maximum value $p=0.592$ is deposited in the Keplerian periodogram. Note that there are the two local optima.}
\end{figure}

Figure~\ref{Fig:powermap} visualises local optima in the $p_{e,T_{0}}$ map at an arbitrary fixed frequency. There are two obvious local optima which means that searching from only one initial value for $T_{0}$ may be not sufficient as one could fail to lead the best local optimum in the $e$-$T_{0}$ plane. This justifies a stepping through $e$ and $T_{0}$. The complexity in the $e$-$T_{0}$ plane, in particular at high eccentricities, finally translates into the Keplerian periodogram. When varying the frequency the landscape and maxima will evolve and the maxima can also switch.

In the given example LS and GLS would give a good initial guess for the best Keplerian period with only a slight frequency shift. But this is not always the case.

One may argue, that the second peak has an equal height suggesting the same significance. On a linear scale it seems so. But the significance is not a linear function of the power. \citet{Cumming08} normalised the Keplerian periodogram as
\begin{equation}
z_{\mathrm{Kep}}(\omega)=\frac{N-5}{4}\frac{\chi_{0}^{2}-\chi^{2}(\omega)}{\chi_{\mathrm{best}}^{2}}=\frac{N-5}{4}\frac{p_{\mathrm{Kep}}(\omega)}{1-p_{\mathrm{Kep}}(\omega_{\mathrm{best}})}\label{eq:z_power}
\end{equation}
analogous to Eq.~(\ref{eq:NormCumming}) and derived the probability distribution
\[
\mathrm{Prob}(z>z_{0})=\left(1+\frac{N-3}{2}\frac{4z_{0}}{N-5}\right)\left(1+\frac{4z_{0}}{N-5}\right)^{-\frac{N-3}{2}}.\]
With this we can calculate that the higher peak which is much closer to 1 has a $10^{-14}$ times lower probability to be due to noise, i.e. it has a $10^{14}$ times higher significance. In Fig.~\ref{Fig:LSvsGLSlogscale} the periodogram power is plotted on a logarithmic scale for $\chi^{2}$ on the right-hand side. The much lower $\chi^{2}$ is another convincing argument for the much higher significance.

\citet{Cumming04} suggested to estimate the FAP for the period search analogous to Eq.~(\ref{eq:FAPapprox}) and the number of independent frequencies again as $M\approx T\Delta f$. This estimation does not take into account the higher variability in the Keplerian periodogram, which depends on the examined eccentricity range, and therefore this FAP is likely to be underestimated.

\begin{figure}
\includegraphics{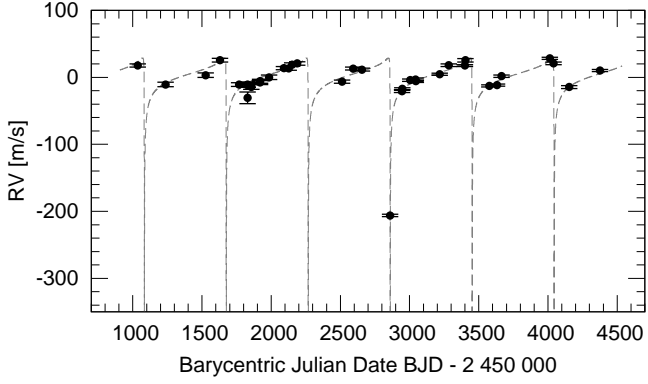}
\caption{\label{Fig:RVHD20782}The radial velocity (RV) time series of HD~20782. The solid line is the best Keplerian orbit fit.}
\end{figure}

\begin{figure}
\includegraphics{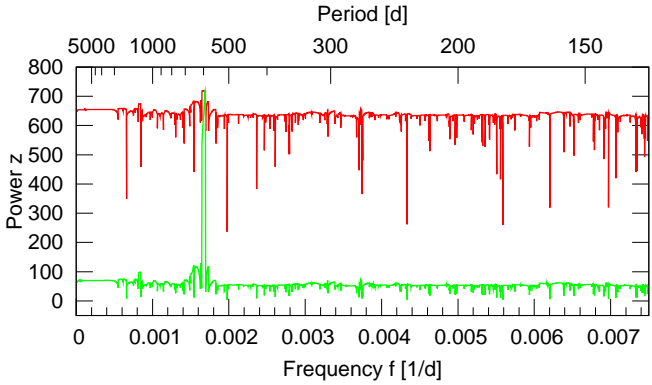}
\caption{\label{Fig:KLS_HD20782}Keplerian periodogram for HD~20782. Both are the same Keplerian periodogram, but the upper one is normalized with the best fit (Eq.~(\ref{eq:z_power})) while the lower one is normalized with the best fit \emph{at each} frequency (Eq.~(\ref{eq:z_power2})). Both have by definition the same maximum value.}
\end{figure}

Another, more extreme example is the planet around HD~20782 discovered by \citet{Jones06}. Figure~\ref{Fig:RVHD20782} shows the RV data for the star taken from \citet{OToole08}. Due to the high eccentricity this is a case where LS and GLS fail to find the right period. However, our algorithm for the Keplerian periodogram find the same solution as the 2DKLS ($P=591.9\,$d, $e=0.97$). The Keplerian periodogram in Fig.~\ref{Fig:KLS_HD20782} indicates this period. This time it is normalised according to Eq.~(\ref{eq:z_power}) and seems to suffer from an overall high noise level (caused by many other eccentric solutions that will fit the one 'outlier'). However, that the period is significant can again be shown just as in the previous example.

For comparison we also show the periodogram with the normalisation by the best fit \emph{at each} frequency \citep{Cumming04}
\begin{align}
z_{\mathrm{Kep}}(\omega) & =\frac{N-5}{4}\frac{\chi_{0}^{2}-\chi^{2}(\omega)}{\chi^{2}(\omega)}\label{eq:z_power2}
\end{align}
which is used in the 2DKLS and reveals the power maximum as impressively as in \citet[Fig.1b]{OToole08}. As \citet{Cumming99} mentioned the choice of the normalisation is a matter of taste; the distribution of maximum power is the same. Finally, keep in mind when comparing Fig.~\ref{Fig:KLS_HD20782} with the 2DKLS in \citet[Fig.1b]{OToole08} which shows a slice at $e=0.97$, that the Keplerian periodogram in Fig.~\ref{Fig:KLS_HD20782} includes all eccentricities ($0\le e\le0.99$). Also the algorithms are different because we also step for $T_{0}$ and simultaneously fit the longitude of periastron $\varpi$.

\section{Conclusions}

Generalised Lomb-Scargle periodogram (GLS), floating-mean periodogram \citep{Cumming99}, date-compensated discrete Fourier transform (DCDFT, \citealt{Ferraz81}), and ``spectral significance'' (SigSpec, \citealt{Reegen07}) at last all mean the same thing: least-squares spectrum for fitting a sinusoid plus a constant. \citet{Cumming99} and \citet{Reegen07} have already shown the advantages of accounting for a varying zero point and therefore we recommend the usage of the generalised Lomb-Scargle periodogram (GLS) for the period analysis of time series. The implementation is easy as there are only a few modifications in the sums of the Lomb-Scargle periodogram.

The GLS can be calculated as conveniently as the Lomb-Scargle periodogram and in a straight forward manner with an analytical solution while programs applying standard routines for fitting sinusoids involve solving a set of linear equations by inverting a 3$\,$x$\,$3 matrix repeated at each frequency. The GLS can be tailored by concentrating the sums in one loop over the data. As already mentioned by \citet{Lomb76} Eq.~(\ref{eq:Periodogram}) (including Eqs.~(\ref{eq:CC})--(\ref{eq:YSal})) should be applied for the numerical work. A fast calculation of the GLS is especially desirable for large samples, large data sets and/or many frequency steps. It also may be helpful to speed up prewhitening procedures \citep[e.g.][]{Reegen07} in case of multifrequency analysis or numerical calculations of the significance of a signal with bootstrap methods or Monte Carlo simulations.

The term generalised Lomb-Scargle periodogram has already been used by \citet{Bretthorst01} for the generalisation to sinusoidal functions of the kind: $y(t)=aZ(t)\cos\omega t+bZ(t)\sin\omega t$ with an arbitrary amplitude modulation $Z(t)$ whose time dependence and all parameters are fully specified (e.g. $Z(t)$ can be an exponential decay). Without repeating the whole procedure given in \ref{sub:Deriv_Periodogram}, \ref{sub:Proof_tau} and Sect.~\ref{sec:GLS} it is just mentioned here that the generalisation to $y(t)=aZ(t)\cos\omega t+bZ(t)\sin\omega t+c$ will result in the same equations with the difference that $Z(t_{i})$ has to be attached to each sine and cosine term in each sum (e.g. $\hat{CC}=\sum w_{i}Z(t_{i})\cos\omega t_{i}\cdot Z(t_{i})\cos\omega t_{i}$).

We presented an algorithm for the application of GLS to the Keplerian periodogram which is the least-squares spectrum for Keplerian orbits. It is an hybrid algorithm that partially applies an analytic solution for linearised parameters and partially steps through non-linear parameters. This has to be distinguished from methods that use the best sine fit as an initial guess. With two examples we have demonstrated that our algorithm for the computation of the Keplerian periodogram is capable to detect (very) eccentric planets in a systematic and nonrandom way.

Apart from this, the least-squares spectrum analysis (the idea goes back to \citealp{Vanicek71}) with more complicated model functions than full sine functions is beyond the scope of this paper (e.g. including linear trends, \citealp{Walker95} or multiple sine functions). For the calculation of such periodograms the employment of the analytical solutions is not essential, but can be faster.

\appendix

\section{Appendix}

\subsection{\label{sub:Deriv_Periodogram}Derivation of the generalised Lomb-Scargle periodogram (GLS)}

The derivation of the generalised Lomb-Scargle periodogram is briefly shown. With the sinusoid plus constant model
\[
y(t)=a\cos\omega t+b\sin\omega t+c\]
the squared difference between the data $y_{i}$ and the model function $y(t)$
\[
\chi^{2}=W\sum w_{i}[y_{i}-y(t_{i})]^{2}\]
is to be minimised. For the minimum $\chi^{2}$ the partial derivatives vanish and therefore:
\begin{align}
0=\partial_{a}\chi^{2} & =2W\sum w_{i}[y_{i}-y(t_{i})]\cos\omega t_{i}\label{eq:Partial_a}\\
0=\partial_{b}\chi^{2} & =2W\sum w_{i}[y_{i}-y(t_{i})]\sin\omega t_{i}\label{eq:Partial_b}\\
0=\partial_{c}\chi^{2} & =2W\sum w_{i}[y_{i}-y(t_{i})]\label{eq:Partial_c}
\end{align}
These conditions for the minimum give three linear equations:
\[
\left[\begin{array}{c}
\hat{YC}\\
\hat{YS}\\
Y\end{array}\right]=\left[\begin{array}{ccc}
\hat{CC} & \hat{CS} & C\\
\hat{CS} & \hat{SS} & S\\
C & S & 1\end{array}\right]\left[\begin{array}{c}
a\\
b\\
c\end{array}\right]\]
where the abbreviations in Eqs.~(\ref{eq:Y})-(\ref{eq:CS}) were applied. Eliminating $c$ in the first two equations with the last equation ($c=Y-aC-bS$) yields:
\[
\left[\begin{array}{c}
\hat{YC}-Y\cdot C\\
\hat{YS}-Y\cdot S\end{array}\right]=\left[\begin{array}{cc}
\hat{CC}-C\cdot C & \hat{CS}-C\cdot S\\
\hat{CS}-C\cdot S & \hat{SS}-S\cdot S\end{array}\right]\left[\begin{array}{c}
a\\
b\end{array}\right]\]
Using again the notations of Eqs.~(\ref{eq:YC})-(\ref{eq:CS}) this can be written as:
\[
\left[\begin{array}{c}
YC\\
YS\end{array}\right]=\left[\begin{array}{cc}
CC & CS\\
CS & SS\end{array}\right]\left[\begin{array}{c}
a\\
b\end{array}\right].\]
So the solution for the parameters $a$ and $b$ is
\begin{align}
a=\frac{YC\cdot SS-YS\cdot CS}{D} &  & \text{and} &  & b=\frac{YS\cdot CC-YC\cdot CS}{D}.\label{eq:Ampl}
\end{align}
The amplitude of the best-fitting sine function at frequency $\omega$ is given by $\sqrt{a^{2}+b^{2}}$. With these solutions the minimum $\chi^{2}$ can be written only in terms of the sums Eqs.~(\ref{eq:YY})-(\ref{eq:CS}) when eliminating the parameters $a$, $b$ and $c$ as shown below. With the conditions for the minimum Eqs.~(\ref{eq:Partial_a})-(\ref{eq:Partial_c}) it can be seen that:
\begin{align*}
\sum w_{i}[y_{i}-y(t_{i})]y(t_{i})= & \quad\, a\sum w_{i}[y_{i}-y(t_{i})]\cos\omega t_{i}\\
 & +b\sum w_{i}[y_{i}-y(t_{i})]\sin\omega t_{i}\\
 & +c\sum w_{i}[y_{i}-y(t_{i})]\\
= & \:0.
\end{align*}
Therefore, the minimum $\chi^{2}$ can be written as:
\begin{align*}
\chi^{2}(\omega)/W & =\sum w_{i}[y_{i}-y(t_{i})]y_{i}-\underbrace{\sum w_{i}[y_{i}-y(t_{i})]y(t_{i})}_{=0}\\
 & =\hat{YY}-a\hat{YC}-b\hat{YS}-cY\\
 & =\hat{YY}-Y\cdot Y-a(\hat{YC}-Y\cdot C)-b(\hat{YS}-Y\cdot S)\\
 & =YY-aYC-bYS
\end{align*}
where in the last step again the definitions of Eqs.~(\ref{eq:YY})-(\ref{eq:YS}) were applied. Finally $a$ and $b$ can be substituted by Eq.~(\ref{eq:Ampl}):
\begin{align*}
\chi^{2}(\omega)/W & =YY-\frac{SS\cdot YC^{2}}{D}-\frac{CC\cdot YS^{2}}{D}+2\frac{CS\cdot YC\cdot YS}{D}.
\end{align*}
When now using the $\chi^{2}$-reduction normalised to unity:
\[
p(\omega)=\frac{\chi_{0}^{2}-\chi^{2}(\omega)}{\chi_{0}^{2}}\]
and the fact that $\chi_{0}^{2}=W\cdot YY$, Eq.~(\ref{eq:Periodogram}) will result.

\subsection{\label{sub:Proof_tau}Verification of Eq.~(\ref{eq:tau})}

Eq.~(\ref{eq:tau}) can be verified with the help of trigonometric addition theorems. For this purpose $CS$ must be formulated. Furthermore, the index $\tau$ and the notation $\varphi=\omega\tau$ will be used:
\begin{align*}
2CS_{\tau}= & \sum w_{i}\sin2(\omega t_{i}-\varphi)\\
 & -2\sum w_{i}\cos(\omega t_{i}-\varphi)\sum w_{i}\sin(\omega t_{i}-\varphi)\\
= & \cos2\varphi\sum w_{i}\sin2\omega t_{i}-\sin2\varphi\sum w_{i}\cos2\omega t_{i}\\
 & -2(C\cos\varphi+S\sin\varphi)(S\cos\varphi-C\sin\varphi)
\end{align*}
Expanding the last term yields
\begin{align*}
2CS_{\tau}= & 2\hat{CS}\cos2\varphi-(\hat{CC}-\hat{SS})\sin2\varphi\\
 & -2\left[C\cdot S(\cos^{2}\varphi-\sin^{2}\varphi)-(C^{2}-S^{2})\sin\varphi\cos\varphi\right]
\end{align*}
and after factoring $\cos2\varphi$ and $\sin2\varphi$:
\begin{align*}
2CS_{\tau}= & 2(\hat{CS}-C\cdot S)\cos2\varphi-\left(\hat{CC}-\hat{SS}-(C^{2}-S^{2})\right)\sin2\varphi\\
= & 2CS\cos2\varphi-(CC-SS)\sin2\varphi
\end{align*}
So for $CS_{\tau}=0$, $\varphi=\omega\tau$ has to be chosen as:
\[
\tan2\omega\tau=\frac{2CS}{CC-SS}\]
By the way, replacing the generalised sums $CC$, $SS$ and $CS$ by the classical ones leads to the original definition for $\hat{\tau}$ in Eq.~(\ref{eq:LStau}):
\begin{align*}
\tan2\omega\hat{\tau} & =\frac{2\hat{CS}}{\hat{CC}-\hat{SS}}=\frac{\sum\sin2\omega t_{i}}{\sum\cos2\omega t_{i}}.
\end{align*}

\bibliographystyle{aa}
\bibliography{paper}

\end{document}